\documentclass[aps,prd,12point,twocolumn,nofootinbib,showpacs,superscriptaddress]{revtex4-2}
\def\theequation{\arabic{section}.\arabic{equation}}
\usepackage{psfrag}
\usepackage{subfigure}
\usepackage{color}
\usepackage{mathrsfs}
\usepackage{graphicx}
\usepackage{amssymb, bm}
\usepackage{amsmath, amsthm}
\usepackage{epstopdf}
\usepackage{hyperref}
\usepackage{enumerate}
\usepackage{longtable}
\usepackage{amsmath}
\usepackage[normalem]{ulem}
\usepackage[title]{appendix}

\newcommand{\be}{\begin{equation}}
\newcommand{\ee}{\end{equation}}
\definecolor{pinegreen}{rgb}{0.0, 0.47, 0.44}

\begin{document}
\def\theequation{\arabic{section}.\arabic{equation}}

\title{Disforming scalar--tensor cosmology}

\author{Valerio Faraoni}
\email[]{vfaraoni@ubishops.ca}
\affiliation{Department of Physics \& Astronomy, Bishop's University, 
2600 College Street, Sherbrooke, Qu\'ebec, Canada J1M~1Z7
}

\author{Carla Zeyn}
\email[]{carla.zeyn@web.de}
\affiliation{Department of Physics, Technical University of Munich, 
Boltzmannstr. 10, 85748 Garching, Germany}
\affiliation{Department of Physics \& Astronomy, Bishop's University, 
2600 College Street, Sherbrooke, Qu\'ebec, Canada J1M~1Z7
}

\begin{abstract} 

Disformal transformations of Friedmann--Lema\^itre--Robertson--Walker and 
Bianchi geometries are analyzed in the context of scalar--tensor gravity. 
Novel aspects discussed explicitly are the $3+1$ splitting, the 
effective fluid 
equivalent of the gravitational scalar, Bianchi models, stealth solutions,  
and de Sitter solutions with non--constant scalar field (which are 
signatures of scalar--tensor gravity). Both pure disformal transformations 
and more general ones are discussed, including those containing 
higher derivatives of the scalar field recently introduced in the 
literature. 

\end{abstract}



\maketitle

\section{Introduction}
\label{sec:1}
\setcounter{equation}{0}

Einstein's theory of gravity, General Relativity (GR), is plagued by 
spacetime singularities, such as those inside black holes or the Big Bang 
singularity in cosmology \cite{Wald:1984rg}. There is hope that quantum 
mechanics and the uncertainty principle on which it is based will some day 
cure these singularities. However, as soon as one tries to 
quantum--correct 
GR, deviations from it are introduced in the form of extra degrees of 
freedom \cite{Brans:1961sx, Bergmann:1968ve, Nordtvedt:1968qs, 
Wagoner:1970vr, Nordtvedt:1970uv}, quadratic terms in the curvature 
invariants appearing in the action \cite{Stelle:1976gc,Stelle:1977ry},  
and higher--order equations of 
motion. For example, the first scenario of early universe inflation, and 
the one currently favoured by observations \cite{Planck:2015sxf}, {\em 
i.e.}, Starobinski inflation \cite{Starobinsky:1980te}, is due to 
quadratic corrections to the Einstein--Hilbert action. The low--energy 
limit of the bosonic string theory, the simplest of string theories, does 
not reproduce GR but gives $\omega=-1$ Brans--Dicke gravity instead 
\cite{Callan:1985ia,Fradkin:1985ys}. Therefore, it is not a matter of if, 
but of where, GR fails and the study of alternative theories of gravity is 
motivated by fundamental physics.

From a completely different point of view, astronomers are also invoking 
modified gravity. The 1998 discovery that the cosmic expansion accelerates 
today left us in the need of an explanation. Dark energy, in the form of a 
fine--tuned cosmological constant, quintessence scalar fields, or a 
plethora of other models, was invoked to explain the cosmic 
acceleration 
(see \cite{AmendolaTsujikawabook} for a review). The standard model of 
cosmology based on GR, the Lambda--Cold Dark Matter ($\Lambda$CDM) model 
is now suffering severe tensions and was never satisfactory because 
approximately 70\% of the energy content of the universe is postulated to 
be dark energy, which was introduced in a completely {\em ad hoc} 
manner and 
whose nature is unknown. Dissatisfied with this state of 
affairs, theorists and astronomers alike have resorted to modifying 
Einstein gravity as an alternative to introducing dark energy 
\cite{Capozziello:2003tk,Carroll:2003wy}. The most popular class of 
theories for this purpose is probably $f(R)$ gravity (see 
\cite{Sotiriou:2008rp, DeFelice:2010aj, Nojiri:2010wj} for reviews), which 
includes the Starobinski action and is ultimately reduced to a 
scalar--tensor gravity.

Scalar--tensor gravity introduces the simplest degree of freedom in 
addition to the two massless spin two modes of GR: a massive propagating 
scalar field which has gravitational nature. The 
prototype of the alternative to 
GR was Brans--Dicke theory, in which the gravitational scalar field $\phi$ 
essentially plays the role of the inverse of the effective gravitational 
coupling strength $G_\mathrm{eff} \simeq 1/\phi $  replacing 
Newton's constant $G$. This effective 
coupling strength becomes a dynamical field sourced by the trace of the 
matter 
stress--energy tensor \cite{Brans:1961sx}. Brans--Dicke theory was later 
generalized to other ``first--generation'' scalar--tensor gravities. In 
the last decade, the search for scalar--tensor theories containing field 
equations with order not higher than second, led to the rediscovery of 
Horndeski gravity \cite{Horndeski}, which has been the subject of an 
extensive literature ({\em e.g.}, \cite{H1,H2,H3, GLPV1, 
GLPV2,Starobinsky:2019xdp} and 
references therein).  More recently, it was found that certain 
higher--order scalar--tensor theories, when subject to a degeneracy 
condition, lead to even more general second order equations of motion. 
These are the so--called Degenerate Higher Order Scalar--Tensor (DHOST) 
theories \cite{DHOST1, DHOST2, DHOST3, DHOST4, DHOST5, DHOST6, 
DHOST7,Langloisetal18, Creminelli:2017sry, 
Creminellietal18, Ezquiaga:2017ekz,Baker:2017hug}, see 
\cite{DHOSTreview1,Langlois:2017mdk} for reviews.

The field equations of DHOST and Horndeski theories are complicated and it 
is very difficult to find non--trivial analytic solutions. 
Many of the known exact solutions of Horndeski and DHOST gravity 
have been obtained by direct integration of the field equations ({\em 
e.g.}, \cite{Minamitsuji:2019tet, deRham:2019gha, Minamitsuji:2019shy}).  
Others, instead, have been obtained by disformal 
transformations of 
the metric tensor $g_{\mu\nu}$, sometimes with the disformal 
transformation reducing to a conformal one. 
 A disformal transformation takes a seed 
metric and scalar field solution $\left( g_{\mu\nu}, \phi \right)$ of a 
Horndeski theory and maps it into a DHOST solution. Here we are interested 
in the disformal transformation of cosmological metrics. The 
transformation of a spatially homogeneous and isotropic 
Friedmann--Lema\^itre--Robertson--Walker (FLRW) metric has been discussed 
in Ref.~\cite{Domenech:2015hka}, see also 
\cite{Adil:2021zxp,Qiu:2020csx} and 
references therein. Here we extend that 
analysis and consider several aspects 
not previously discussed in relation with disformal transformations, 
including the explicit $3+1$ splitting, Bianchi cosmologies, 
and the 
effective fluid 
equivalent of the scalar degree of freedom $\phi$. We also discuss 
certain  exact solutions 
typical of scalar--tensor gravity which are impossible in GR, namely 
stealth solutions, as well as de Sitter solutions with 
non--constant scalar field which, in a sense explained 
below, generalize stealth solutions.

There is further motivation for studying the disformal 
transformation of cosmological spacetimes. Black holes interacting with 
their environment are dynamical and their horizons are not event horizons, 
but apparent horizons isntead, which makes them much more complicated from 
the point of view of black hole mechanics and thermodynamics 
\cite{Faraoni:2015ula, 
Hollands:2024vbe}.  Similarly, dynamical cosmological horizons have 
thermodynamics that is far from trivial \cite{Faraoni:2015ula}. One way to 
make black holes dynamical is to embed them in a non--static cosmological 
``background''. This possibility is of great astrophysical interest since 
the interaction of black holes with the FLRW space in which they are 
embedded, over cosmological time scales, has been tentatively reported in 
recent observations \cite{Farrah:2023opk,Croker:2022vdq}. If supermassive 
black holes at the centres of galaxies are taken to be non--singular 
objects with an extended de Sitter core, as in most models of regular 
black holes, then the possiblity arises that dark energy could be 
effectively segregated inside these black hole horizons 
\cite{Farrah:2023opk, Croker:2022vdq, Croker:2021duf, 
Croker:2020plg,Croker:2019kje, Croker:2019mup}.  Aside from its 
cosmological implications, the possibility of cosmological coupling 
tentatively reported in \cite{Farrah:2023opk} has already been the subject 
of a lively theoretical and observational debate in the literature 
\cite{Mottola:2023jxl, Rodriguez:2023gaa, Lei:2023mke, Amendola:2023ays, 
Garcia-Bellido:2023exi, Avelino:2023rac, Dahal:2023suw, Wang:2023aqe, 
Casadio:2023ymt, Mistele:2023fds, Karakasis:2023hni, Andrae:2023wge, 
Cadoni:2023lum, Carleo:2023qxu, Gao:2023keg, Cadoni:2023lqe, 
Mlinar:2023fkk, Lacy:2023kbb, Dahal:2023hzo, Binici:2024smk, 
Wagner:2023axm}. Problem is, the theory behind this cosmological coupling 
of black holes is underdeveloped and exact solutions of the relevant field 
equations could help understanding the basic physical principles behind 
this phenomenon. The scarcity of relevant solutions in GR 
\cite{Faraoni:2015ula} prompts the search for new solutions in more 
general scalar--tensor and Horndeski theories. The easiest way to generate 
new solutions is by using disformal transformations of GR ``seeds''. The 
first step in this program consists of understanding the transformation 
properties under disformal transformations of the FLRW or Bianchi 
``backgrounds'' in which such black holes are embedded.  The present work 
addresses this first step.

We adopt the notations of Ref.~\cite{Wald:1984rg}. The metric signature is 
${-}{+}{+}{+}$, units are used in which the speed of light $c$ and 
Newton's constant $G$ are unity, and $\Box \equiv g^{\mu\nu}\nabla_{\mu} 
\nabla_{\nu}$ is the curved space d'Alembert operator. Greek indices run 
from~0 to~3 and Latin indices from~1 to~3.

\section{Disformal transformations}
\label{sec:2}
\setcounter{equation}{0}

The general form of a disformal transformation with first order 
derivatives of the scalar field \cite{Bekenstein:1992pj} 
is \be
g_{\mu\nu} \rightarrow \bar{g}_{\mu\nu} = \Omega^2 \left (\phi, X \right) 
g_{\mu\nu} + 
F(\phi, X) \nabla_{\mu}\phi\nabla_{\nu} \phi \,,
\ee
where 
\be
X \equiv  -\frac{1}{2} \, g^{\mu\nu}\nabla_{\mu}\phi \nabla_{\nu} \phi
\ee
and $\phi$ denotes the scalar field (see \cite{Alinea:2020sei} for a 
summary of the transformation properties of various geometrical quantities 
under disformal transformations). In the case $\Omega = 1$, one obtains a 
{\it pure disformal transformation}, to which we restrict for most of this 
paper until Sec.~\ref{sec:8}:
\be
g_{\mu\nu} \rightarrow \bar{g}_{\mu\nu} = g_{\mu\nu} + F(\phi, X) 
\nabla_{\mu}\phi\nabla_{\nu} \phi  \,.\label{puredisformal}
\ee
Assuming that $\nabla^\mu\phi$ is timelike, one can always use the 
``uniform--$\phi$ 
slicing'' of spacetime in which $\phi = \phi(t)$, where $t$ is the time 
coordinate. In this gauge, the line 
element assumes the familiar form 
\begin{eqnarray}
ds^2 &=& g_{\mu\nu} dx^{\mu}dx^{\nu} \nonumber\\
&&\nonumber\\
&=& -N^2dt^2 + g_{ij}\left( 
dx^i + 
N^idt \right) \left( dx^j + N^jdt \right) \,,
\end{eqnarray}
where $N$ and $N^i$ are the lapse and the shift vector, 
respectively. 
The disformed line element is 
\begin{eqnarray}
   d\bar{s}^2 &=& \bar{g}_{\mu\nu} dx^{\mu}dx^{\nu} \nonumber\\
&&\nonumber\\
& = &-\bar{N}^2 dt^2  + \bar{g}_{ij} \left( dx^i + \bar{N}^i dt \right) 
\left( dx^j + \bar{N}^jdt \right)\,, \nonumber\\
&& \label{questa}
\end{eqnarray}
where $\bar{N}^2 = N^2\alpha^2$, $\bar{N}^i = N^i$, $\bar{g}_{ij}  
= g_{ij}$ in the disformed (or barred) world, and \cite{Domenech:2015hka} 
\be
\alpha^2 \equiv  1-2F(\phi, X)X \,.
\ee 
In order to preserve the metric signature, it must be $F < 1/2 $ 
everywhere 
throughout the spacetime manifold, which we assume in the following. 

Equation~(\ref{questa}) is easily proved. Using the 
notation $\dot{\phi} \equiv d\phi/dt$, we have
\begin{eqnarray}  
d\bar{s}^2 &=& \left( g_{\mu\nu} + F\nabla_{\mu}\phi\nabla_{\nu}\phi 
\right) dx^{\mu} dx^{\nu} \nonumber \\ 
&&\nonumber\\ 
& = & ds^2 + F(\phi,X)\left( \nabla_{\mu}\phi \, dx^{\mu} \right)^2 
\nonumber 
\\ 
&&\nonumber\\ 
& = & -N^2dt^2 + g_{ij} \left( dx^i + N^idt \right) \left( dx^j + N^jdt 
\right) + F \dot{\phi}^2dt^2 \nonumber \\ 
&&\nonumber\\ 
& = & -\left( N^2-F \dot{\phi}^2 \right)dt^2   + g_{ij} \left( dx^i + 
N^idt \right) \left( dx^j + N^jdt \right) \,.\nonumber\\
&&\label{new27}
\end{eqnarray}
Since
\begin{equation}
 X = -\frac{g^{00}}{2} \, \dot{\phi}^2 = \frac{\dot\phi^2}{2N^2} \,,
\end{equation}
it is 
\be
N^2 - F\left(\phi, X\right) \dot\phi^2 = N^2\left[ 1-2F(\phi, X)X\right] 
\equiv N^2 \alpha^2 \,.
\ee
From
\begin{equation}
    d\bar{s}^2 = -(N^2\alpha^2)dt^2 + g_{ij}(dx^i+N^idt)(dx^j + N^j dt)
\end{equation}
it then follows that 
\be
\bar{N}^2 = N^2\alpha^2\,, \quad \bar{g}_{ij} = g_{ij}\,, \quad \bar{N}^i 
= N^i\,.
\ee

The inverse of the disformed metric $\bar{g}_{\mu\nu}$  is 
\cite{Domenech:2015hka, Alinea:2020sei}
\be
\bar{g}^{\mu\nu} = g^{\mu\nu} - \frac{F}{\alpha^2} 
\, \nabla^{\mu} \phi \nabla^{\nu} \phi \,:
\ee
to wit, 
\begin{eqnarray}   
\bar{g}^{\mu\nu} \bar{g}_{\nu\gamma} 
&  = & \left( g^{\mu\nu} - \frac{F}{\alpha^2} \, \nabla^{\mu} \phi 
\nabla^{\nu}  \phi\right) \left( g_{\nu\gamma} + F \, \nabla_{\nu} \phi 
\nabla_{\gamma} \phi \right) 
\nonumber\\ 
&&\nonumber\\ 
& = & g^{\mu\nu} g_{\nu\gamma} + 
F \left( g^{\mu\nu} \nabla_{\nu} \phi\nabla_{\gamma}\phi 
-\frac{1}{\alpha^2} \, g_{\nu\gamma}\nabla^{\mu} \phi\nabla^{\nu} \phi 
\right) \nonumber\\ 
&&\nonumber\\
& \, & -  \frac{F^2}{\alpha^2} 
\left(\nabla_{\nu} \phi\nabla^{\nu} \phi \right) 
\nabla^{\mu} \phi\nabla_{\gamma} \phi 
\nonumber\\ &&\nonumber\\ 
& = & \delta^{\mu}_{\gamma} + F(\phi,X)\left(\nabla^{\mu} 
\phi\nabla_{\gamma}\phi - 
\frac{\nabla^{\mu} \phi\nabla_{\gamma} \phi}{1-2F(\phi,X)X}\right) 
\nonumber\\
&&\nonumber\\
&\, & - \frac{F^2(\phi,X)}{1-2F(\phi,X)X} \left( 
\nabla_{\nu} \phi\nabla^{\nu} \phi \right)\nabla^{\mu} 
\phi\nabla_{\gamma} \phi \nonumber\\
&&\nonumber\\ 
& = & \delta^{\mu}_{\gamma} + \frac{F\left( 1-2FX-1 \right) - 
F^2(-2X)}{1-2FX} \, \nabla^{\mu} \phi\nabla_{\gamma} \phi \nonumber\\
&&\nonumber\\
& = & \delta^{\mu}_{\gamma} \,.\label{new3}
\end{eqnarray}
Let us proceed to adapt these transformations to FLRW geometries.

\section{FLRW universe and disformal transformations}
\label{sec:3}
\setcounter{equation}{0}

The FLRW line element in comoving coordinates $\left( t,r, \vartheta, 
\varphi \right)$ is 
\begin{eqnarray}  
ds^2 &=& -dt^2 + a^2(t) \left( \frac{dr^2}{1-K r^2} + 
r^2d\Omega_{(2)}^2 \right) \nonumber\\
&& \nonumber\\
& \equiv & -dt^2 + g_{ij} dx^idx^j \,,
\end{eqnarray}
where the constant $K$ is the curvature index, $N = 1$, $N^i = 0$, and 
$d\Omega_{(2)}^2 
\equiv  d\vartheta^2 + \sin^2 \vartheta \, d\varphi^2$ is the line 
element on the unit 2--sphere. The pure disformal 
transformation~(\ref{puredisformal}) changes the 
FLRW line element into 
\be
d\bar{s}^2 = -\left( 1-2FX \right)dt^2 + g_{ij}dx^idx^j \,.
\ee
The uniform--$\phi$ slicing corresponds to the comoving FLRW slicing in 
which 
$\phi = \phi(t)$. In this gauge  $X = \dot\phi^2 /2 $ and 
\be
d\bar{s}^2 = - \left( 1-\dot{\phi}^2 F \right)dt^2 + g_{ij}dx^i dx^j
\ee
after the disformal transformation. By redefining the time coordinate 
according to 
\be
dt^2 \rightarrow d\tau^2 \equiv  \left[ 1-\dot{\phi}^2(t) F(t)\right] dt^2 
\,,
\ee
or
\be
\tau(t) = \int dt \sqrt{1-\dot{\phi}^2(t)F(t)} \,,\label{tau}
\ee
we rewrite the disformed line element as  
\be
d\bar{s}^2 = -d\tau^2 + g_{ij} dx^i dx^j 
\ee
and $ \bar{\phi} (\tau)=\phi\left( t(\tau) \right)$. 
Therefore, a pure disformal transformation of a FLRW metric 
written in the 
uniform--$\phi$ gauge always generates another FLRW 
geometry, in addition to preserving the 
uniform--$\phi$ slicing.\footnote{If 
$ \left( g_{\mu\nu}, \phi \right)$ is a GR solution with minimally coupled 
scalar $\phi$, then $\left( \bar{g}_{\mu\nu},\phi \right)$ is a solution of 
``class Ia'' DHOST gravity \cite{BenAchour:2016cay}.} 
In essence, a pure disformal transformation of a FLRW metric is equivalent 
to a rescaling of the comoving time \cite{Domenech:2015hka}. For 
example, for spatially flat FLRW metrics   
\be
ds^2 = -dt^2 + a^2(t) \left( dx^2 + dy^2 + dz^2 \right) \equiv  -dt^2 + 
a^2(t) d\vec{x}^{\,2} \,,
\ee
for which
\begin{eqnarray}  
d \bar{s}^2 &=& (g_{\mu\nu} + 
F \, \nabla_{\mu}\phi\nabla_{\nu}\phi)dx^{\mu}dx^{\nu} \nonumber\\
&& \nonumber\\
& = & g_{\mu\nu} dx^{\mu} dx^{\nu} + F\dot{\phi}^2 dt^2 \nonumber\\
&& \nonumber\\
& = & -\left( 1-F\dot{\phi}^2 \right)dt^2 + a^2(t)d\vec{x}^{\, 2} 
\nonumber\\
&& \nonumber\\
& = & -d\tau^2 + \bar{a}^2(\tau) d\vec{x}^{\, 2} \,.
\end{eqnarray}
The pure disformal transformation yields another 
spatially flat FLRW line 
element with comoving time~(\ref{tau}) and scale factor 
$\bar{a}(\tau)=a\left( t(\tau)\right)$.

Both in GR with a minimally coupled scalar field $\phi$ 
\cite{EllisWainwright1997,Coley:2003mj,Faraoni:2012bf,Remmen:2013eja} and in 
``first--generation'' scalar--tensor gravity \cite{Faraoni:2005vc} 
(including $f(R)$ gravity \cite{deSouza:2007zpn}), spatially flat 
 FLRW universes are described by the dynamical variables $H 
\equiv 
\dot{a}/a$ and $\phi$. That is, for these $K=0$ FLRW universes,  the scale 
factor $a(t)$ enters the 
Einstein--Friedmann equations only through the  Hubble function 
$\dot{a}/a$, which is a cosmological observable. 
Hence, the phase space is the $\left( H, \phi, \dot{\phi} \right)$ space 
but the Friedmann equation
\be
H^2= \frac{8\pi }{3} \, \rho^{(\phi)}
\ee
constitutes a first order constraint on the dynamics. As a consequence,  
the region of the phase space 
accessible to the orbits of the solutions is a 2--dimensional subset of 
this 3--dimensional space. Effectively, these orbits move on a curved 
2--dimensional subset of the 3--space $\left( H, \phi, \dot{\phi} 
\right)$, which may consist of multiple sheets and possibly have 
``holes'' inaccessible to these orbits, as explained in 
Refs.~\cite{Faraoni:2005vc, deSouza:2007zpn}. This phase space structure 
extends 
to spatially flat FLRW cosmology in Horndeski gravity 
\cite{Edwards:2016xzx}.

In  both GR and scalar--tensor 
gravity, if fixed 
points exist, with this choice of dynamical variables they  are 
unavoidably de Sitter spaces with 
\be
\left( H, \phi \right) =  \left( H_0=\mbox{const.},
\phi_0=\mbox{const.} \right) \,.
\ee
The values of $H_0$ and $\phi_0$ are related by the 
field equations 
\cite{Faraoni:2005vc, deSouza:2007zpn}. 
A disformal transformation of a $K=0$ FLRW metric  maps these fixed 
points into FLRW universes with 
\be
\bar{g}_{\mu\nu} = g_{\mu\nu}^\mathrm{(FLRW)}+ F \, \nabla_{\mu} \phi 
\nabla_{\nu} \phi \,, 
\quad 
\bar{\phi}(\tau) =\phi\left(t(\tau) \right) \,,
\ee
where $d\tau=\sqrt{ 1-2FX} \, dt$ but, since for fixed points 
$\nabla_{\alpha} 
\phi=\nabla_{\alpha} \phi_0=0$, it is simply $\bar{g}_{\mu\nu} = 
g_{\mu\nu}^\mathrm{(FLRW)}$. Therefore,  
\begin{center} {\em a pure disformal 
transformation~(\ref{puredisformal}) maps de Sitter fixed points into de 
Sitter fixed points of the phase space of $K=0$ FLRW cosmology}.
\end{center}

de Sitter spaces with non--constant scalar fields do not exist in GR with 
minimally coupled scalar, but are a signature of scalar--tensor gravity. 
They are not mapped into de Sitter spaces, as discussed in 
Sec.~\ref{sec:7}.

\section{Effective imperfect fluid of scalar--tensor gravity and disformal 
transformations}
\label{sec:4}
\setcounter{equation}{0}

It is well known that the scalar field $\phi$ of scalar--tensor 
gravity can be seen as an  imperfect fluid when its gradient $\nabla^{\mu} 
\phi$ is timelike  and future--oriented, {\em i.e.}, 
$t^{\alpha}\nabla_{\alpha} \phi <  0$, where $t^{\mu}= \left( 
\partial/\partial t \right)^{\mu} $ 
is the time 
direction of observers comoving with this effective fluid 
\cite{Pimentel89, Faraoni:2018qdr, Quiros:2019gai, 
Giusti:2021sku}.  As is customary, we write the 
vacuum field equations of scalar--tensor gravity in 
the form of effective Einstein equations,  
\be
G_{\mu\nu} = T_{\mu\nu}^{(\phi)} \,;
\ee
then, we assume $\nabla^{\mu} \phi$ to be 
timelike and future--oriented. One can then   
define the effective fluid four--velocity 
\be
u^{\mu} = \frac{\nabla^{\mu} \phi}{\sqrt{2X}} \,,
\ee
which satisfies the usual normalization condition for 
timelike fluids $u^{\alpha} u_{\alpha} = -1$. The effective 
stress--energy 
tensor of the gravitational scalar field  
$\phi$ has the form
\be
T^{(\phi)}_{\mu\nu} = \rho u_{\mu} u_{\nu} + P h_{\mu\nu} + \pi_{\mu\nu} 
+ q_{\mu} u_{\nu}  + q_{\nu} u_{\mu}  \,,
\ee
where 
\be
\rho = T_{\mu\nu} u^{\mu} u^{\nu} 
\ee
is the effective energy density, 
\be
P = \frac{1}{3} \, h^{\mu\nu} T_{\mu\nu}
\ee
is the effective isotropic pressure,  
\be
\pi_{\mu\nu} = T_{\gamma\delta} {h_{\mu}}^{\gamma} {h_{\nu}}^{\delta} - P 
\, h_{\mu\nu}
\ee
is the effective anisotropic stress tensor, and 
\be
q_{\alpha} = - T_{\gamma\delta} \, u^{\gamma} {h_{\alpha}}^{\delta} 
\ee
 is the effective heat flux density  \cite{Pimentel89, Faraoni:2018qdr, 
Quiros:2019gai, Giusti:2021sku}. Here, 
\be
h_{\mu\nu} = g_{\mu\nu} + u_{\mu} u_{\nu}
\ee
is the Riemannian metric in the 3--space orthogonal to $u^{\alpha}$  
and the pressure $P$ is the sum of non--viscous and viscous 
contributions, 
\be
P = P_0 + P_\mathrm{vis} \,.
\ee
$h_{\mu\nu}$, $\pi_{\mu\nu}$, and $q_{\mu}$ 
are purely spatial:
\be
h_{\mu\nu} u^{\mu} = h_{\mu\nu} u^{\nu} = \pi_{\mu\nu} u^{\mu} = \pi_{\mu\nu} 
u^{\nu} = q_{\mu} u^{\mu} = 0 \,.
\ee

Let us adapt this effective fluid analogy to FLRW geometries.  

Since a FLRW universe is spatially homogeneous and isotropic, the 
shear tensor ${\pi^i}_j$ and the heat flux density $q^i$ (which would 
introduce a preferred spatial direction if it were non--vanishing), are 
identically zero and  the 
only dissipative quantity that can remain is the viscous pressure 
$P_\mathrm{vis}$. In Eckart's thermodynamics, viscous pressure arises 
because of bulk viscosity, according to the constitutive 
relation $P_\mathrm{vis} = -\zeta \, \nabla_{\rho} u^{\rho}$ 
\cite{Eckart:1940te,Ganguly:2021pke}, where $\zeta$ is a bulk viscosity 
coefficient. When one applies Eckart's first order thermodynamics to the 
effective fluid of the scalar field $\phi$,  this  
relation is satisfied in the 
context of ``old'' scalar--tensor gravity \cite{Brans:1961sx, 
Bergmann:1968ve, Nordtvedt:1968qs, Wagoner:1970vr, Nordtvedt:1970uv}), 
but is usually invalid in more general Horndeski gravity 
\cite{Giusti:2021sku,Miranda:2022wkz}.

Next, one wonders how the effective $\phi$--fluid quantities transform 
under  disformal transformations, but does this question make sense? In 
general spacetimes, it does not, as explained in the following. 
In general, solutions of a 
theory of gravity 
(say, ``theory~A'') are mapped into solutions of a new theory 
(say, ``theory~B'') by 
a disformal transformation. Then, the effective 
stress--energy tensor of $\phi$ 
obtained by recasting the field equations of theory~A as effective 
Einstein equations, will have a different form in the new theory~B. 
For 
example, solutions of GR with a minimally coupled scalar field, or of 
``first--generation'' scalar--tensor gravity (theory~A), are mapped into 
solutions of a Horndeski or a DHOST theory (theory~B) 
\cite{BenAchour:2016cay}. In this case, it 
no longer makes sense to 
consider the original $T_{\mu\nu}^{(\phi)}$, which is replaced by a more 
complicated expression \cite{BenAchour:2016cay}. However, if one restricts 
one's attention to FLRW metrics in the uniform--$\phi$ gauge, the 
disformal transformation amounts to a mere time rescaling and it still 
makes sense to consider the same $T_{\mu\nu}^{(\phi)}$.  
Here we restrict to 
the ``old'' scalar--tensor theories for simplicity, 
however, the extension of this analysis to the effective 
dissipative 
stress--energy tensor of Horndeski gravity is straightforward. The 
derivation of this  
effective stress--energy tensor of $\phi$ is rather laborious and 
is performed in Refs.~\cite{Quiros:2019gai,Giusti:2021sku}. A similar 
situation occurs with 
purely conformal transformations. In this case, starting 
with a solution $\left( g_{\mu\nu}, 
\phi \right)$ of the coupled Einstein--Klein--Gordon equations, a 
conformally transformed metric $\tilde{g}_{\mu\nu}=\Omega^2(\phi) 
g_{\mu\nu}$ is 
no longer a solution of the Einstein equations with the same matter 
content.\footnote{Indeed, the new metric and scalar field obtained 
in this way are solutions of a different theory of gravity. 
Given an electrovacuum solution 
$g_{\mu\nu}$ of GR and an 
arbitrary scalar field $\psi>0$, one can always perform a conformal 
transformation so that $\tilde{g}_{\mu\nu}= g_{\mu\nu}/ \sqrt{\psi}$ is a 
solution 
of an $\omega=-3/2$ Brans--Dicke theory with that scalar $\psi$ as 
its Brans-Dicke scalar \cite{Hammad:2018hhv}. This 
theory is 
pathological since $\psi$ is not dynamical.} 
However, {\em for FLRW metrics and $\phi=\phi(t)$ dependent 
only on time}, the conformal 
transformation amounts again to a rescaling of the comoving time and a 
perfect fluid is mapped again into a perfect fluid with the same 
equation of 
state \cite{Faraoni:2004pi}.

With this {\em caveat}, let us proceed to derive the transformation of 
the various effective fluid quantities in FLRW spaces, which 
are the subject of interest in this article. The 
contravariant four--velocity is 
\be
\bar{u}^{\mu} = \left(\frac{\partial}{\partial\tau}\right)^{\mu} = 
\frac{1}{\alpha} \left(\frac{\partial}{\partial t}\right)^{\mu} = 
\frac{u^{\mu} }{ \alpha} \,.
\ee
The covariant four--velocity is computed as
\begin{eqnarray}  
\bar{u}_{\mu} &=&  \bar{g}_{\mu\nu} \bar{u}^{\nu} = \left( g_{\mu\nu} + 
F\nabla_{\mu}\phi\nabla_{\nu} \phi \right) \frac{u^{\nu}}{\alpha} 
\nonumber\\
&& \nonumber\\
& = & g_{\mu\nu} \, \frac{u^{\nu}}{\alpha} + \frac{F}{\alpha} \, u_{\mu} 
u_{\nu} 
\left( -\nabla^{\gamma} \phi\nabla_{\gamma} \phi \right) u^{\nu} \nonumber 
\\
&& \nonumber\\
& = & \frac{u_{\mu}}{\alpha} + \frac{F}{\alpha} 
\left( \nabla^{\gamma} \phi\nabla_{\gamma} \phi \right)u_{\mu} = 
\frac{u_{\mu} \left( 1-2FX \right)}{\alpha} 
\nonumber\\ && \nonumber\\
& = & \alpha u_{\mu} \,. \label{eq:covariant4-velocity}
\end{eqnarray}

Using the 3+1 splittings
\be
g_{\mu\nu} = -u_{\mu} u_{\nu} + h_{\mu\nu} \,,\quad 
\bar{g}_{\mu\nu} = -\bar{u}_{\mu} \bar{u}_{\nu} + \bar{h}_{\mu\nu} \,,
\ee
the disformal transformation~(\ref{puredisformal}) gives
\begin{eqnarray}  
 \bar{g}_{\mu\nu} &=& g_{\mu\nu} + F \, \nabla_{\mu}\phi\nabla_{\nu} \phi 
\nonumber\\
&& \nonumber\\
& =& -u_{\mu} u_{\nu} + h_{\mu\nu} + 2XF u_{\mu} u_{\nu} \nonumber\\
&& \nonumber\\
& = & -(1-2FX)u_{\mu} u_{\nu} + h_{\mu\nu} \nonumber\\
&& \nonumber\\
& = & -\alpha^2 \, \frac{\bar{u}_{\mu} }{\alpha} \, 
\frac{\bar{u}_{\nu} }{\alpha} + 
h_{\mu\nu} \nonumber\\
&& \nonumber\\
& = & -\bar{u}_{\mu} \bar{u}_{\nu} + h_{\mu\nu} \,;
\end{eqnarray}
the spatial 3--metric is not affected by the disformal 
transformation~(\ref{puredisformal}). 

Next, we determine how $T_{\mu\nu}^{(\phi)}$ transforms under pure 
disformal transformations:
\begin{eqnarray}  
T_{\mu\nu}^{(\phi)} &=& \rho u_{\mu} u_{\nu} + P h_{\mu\nu} \nonumber\\
&& \nonumber\\
& = & \frac{\rho}{\alpha^2} \, \bar{u}_{\mu} \bar{u}_{\nu} + 
P\bar{h}_{\mu\nu}  \nonumber\\
&& \nonumber\\
&\equiv & \bar{\rho} \, \bar{u}_{\mu} \bar{u}_{\nu} + \bar{P} \, 
\bar{h}_{\mu\nu} = \bar{T}_{\mu\nu}^{(\phi)} \,,
\end{eqnarray}
where  the individual fluid quantities transform as
\be
  \bar{\rho} = \frac{\rho}{\alpha^2} = \frac{\rho}{1-2FX} \,, \quad 
\bar{P} = P \,. \label{queste}
\ee
Clearly, the scaling of the energy density with $\alpha$ is a consequence 
solely of the scaling of time, which appears twice because $\bar{\rho} = 
T_{\mu\nu}^{(\phi)} \bar{u}^{\mu} \bar{u}^{\nu} $ is quadratic 
in the four--velocity, which transforms according to 
Eq.~(\ref{eq:covariant4-velocity}). This scaling of 
$u_\alpha$  is 
responsible for the factor 
$\alpha^2$. Purely spatial quantities,  including the pressure $P$, are 
left 
unchanged by time rescalings. Equations~(\ref{queste}) are analogous to 
the well--known transformation relations for energy 
density and pressure under a purely conformal transformation of a perfect 
fluid in FLRW spaces 
\be
g_{\mu\nu} \rightarrow \Tilde{g}_{\mu\nu} = \Omega^2 (\phi) g_{\mu\nu} 
\,.
\ee
These transformation relations are $\tilde{P} = 
\Omega^4 P$, ~$\tilde{\rho} = \Omega^4 \rho$  \cite{Faraoni:2004pi}. 

The effective equation of state parameter of the $\phi$--fluid in the 
barred world is
\be
\bar{w} \equiv \frac{\bar{P}}{\bar{\rho}}  =  \alpha^2 \, \frac{P}{\rho} 
\equiv  
\alpha^2 \, w \,.
\ee
Therefore, $\bar{w}$ has the same sign of $w$ and scales only because of 
the scaling of $\rho$ with $\alpha$. 
If $\alpha>1$ a quintessence scalar field, which by definition has  
$-1<w<-1/3$,  can be mapped into an effective phantom field 
which has instead an equation of state parameter $\bar{w}<-1$.

To summarize, we have
\begin{eqnarray}
&& \bar{u}^{\mu} = \frac{u^{\mu}}{\alpha} \,, \quad u^{\mu} =\alpha \, 
\bar{u}^{\mu} \,,\\
&&\nonumber\\
&& \bar{u}_{\mu} = \alpha \, u_{\mu} \,, \quad u_{\mu} = 
\frac{\bar{u}_{\mu}}{\alpha}\,,\\
&&\nonumber\\ 
&& \bar{h}_{\mu\nu} = h_{\mu\nu} \,,\\
&&\nonumber\\
&& \bar{\rho} = \frac{\rho}{\alpha^2} = \frac{\rho}{1-2FX} \,, \quad 
\bar{P} = P \,. 
\end{eqnarray}

As an example, consider a free, minimally coupled  scalar field in 
GR (which is well--known to behave as a stiff fluid 
\cite{Faraoni:2021opj}) in a spatially flat FLRW universe. The effective 
energy density and isotropic pressure of this $\phi$--fluid 
\be
\rho = \frac{ \dot{\phi}^2}{2} +V \,, \quad
P = \frac{ \dot{\phi}^2}{2} -V 
\ee
coincide when $V(\phi)=0$, giving the effective equation of state 
parameter $w \equiv 
P/\rho=1$ and \be \rho=\frac{\rho_0}{a^6} \,, \quad a(t)=a_0 t^{1/3} \,,
\ee
where $\rho_0$ and $a_0$ are constants. The Klein--Gordon equation  
satisfied by this minimally coupled scalar field  
\be
\Box \phi=- \left( \ddot{\phi}+3H\dot{\phi} \right)=0
\ee
has the solution $ \phi(t)=\phi_0 \ln\left( t / t_0 \right) +\phi_1 
$, where $\phi_{0,1}$ and $t_0$ are constants. 
Moreover, we have
\be
X= \frac{\phi_0^2}{2 t^2} \,, \quad  F(\phi,X) = F(t) \,, \quad
\alpha^2 =1-\frac{\phi_0^2 \, F}{t^2} 
\ee
and
\be
\bar{w}(t)= \alpha^2 w =1-\frac{\phi_0^2 \, F(t)}{t^2} \,.
\ee
For example, the choice
\be
F(X)= \frac{F_0}{X} =\frac{2F_0 t^2}{\phi_0^2} \,,
\ee
where $F_0$ is a constant, gives another constant equation of state 
parameter $\bar{w}= 1-2F_0 \neq w =1$. In GR, a disformal 
transformation can create any constant equation of state with $\bar{w}>0$ 
starting from a stiff fluid with $w=1$.

\section{Bianchi universes and disformal transformations}
\label{sec:5}
\setcounter{equation}{0}

Although not mentioned explicitly in the literature, the previous 
discussion 
applies almost without changes to spatially homogenous but anisotropic 
Bianchi cosmologies \cite{EMMacC}. The line element can be written in the 
form
\be
ds^2 = -dt^2 + \gamma_{ij} dx^i dx^j 
\ee
in the uniform--$\phi$ gauge, which is possible because Bianchi 
cosmological models are spatially homogenous.  
In this gauge $\phi = \phi (t)$, we have 
\begin{eqnarray}  
& &  d\bar{s}^2 = \bar{g}_{\mu\nu} dx^{\mu} dx^{\nu} = (g_{\mu\nu} + 
F\nabla_{\mu}\phi\nabla_{\nu}\phi)dx^{\mu} dx^{\nu} \nonumber\\
&& \nonumber\\
& & = ds^2 + F\dot{\phi}^2 dt^2 = -(1-F\dot{\phi}^2)dt^2 + \gamma_{ij} 
dx^i dx^j \nonumber\\
&&
\end{eqnarray}
where, again,  $F \left(\phi(t), X(t) \right) = F(t)$, so 
\be
d\bar{s}^2 = -d\tau^2 + \gamma_{ij} dx^i dx^j \,.
\ee
As for FLRW geometries, a disformal transformation~(\ref{puredisformal}) 
of a Bianchi metric is equivalent to a time rescaling, produces another 
Bianchi metric, and preserves the uniform--$\phi$ gauge.

As an example, consider the spatially flat Bianchi~I geometry with line 
element
\be
ds^2_{(I)} = -dt^2 + a_1^2 (t) \, dx^2 + a_2^2(t) \, dy^2 + a_3^2 (t) 
\, dz^2 
\ee
in comoving Cartesian coordinates $\left( t,x,y,z \right)$, where $a_i(t)$ 
($i=1,2,3$) are the scale factors corresponding to the three spatial 
directions, and $ \phi = \phi(t)$. The disformal transformation of the 
metric~(\ref{puredisformal}) then yields
\begin{eqnarray}  
d\bar{s}^2 &=& 
 -(1-F\dot{\phi}^2) dt^2 + a_1^2(t) \, dx^2 + a_2^2(t) \, dy^2 + a_3^2(t) 
\, dz^2\nonumber\\
&& \nonumber\\
& = & -d\tau^2 + \bar{a}_1^2(\tau) \, dx^2 + \bar{a}_2^2(\tau) \, dy^2 + 
\bar{a}_3^2(\tau) \, dz^2 \nonumber\\
\end{eqnarray}
with $\tau(t) = \int dt \sqrt{1-F\dot{\phi}^2} $ and $ 
\bar{a}_i(\tau)=a_i\left( t(\tau)\right)$. Since $F$ and 
$\dot{\phi}$ depend only on $t$, the new time coordinate 
$\tau(t)$ is well--defined, because  $d\tau = \sqrt{1-F\dot{\phi}^2}\, dt$ 
is an exact differential.

\section{Stealth solutions of scalar--tensor gravity}
\label{sec:6}
\setcounter{equation}{0}

Stealth solutions \cite{Ayon-Beato:2004nzi, Ayon-Beato:2005yoq,  
Faraoni:2010mj, Ayon-Beato:2013bsa, Caldarelli:2013gqa, 
BravoGaete:2013djh, BravoGaete:2013acu, Hassaine:2013cma, 
Bravo-Gaete:2014haa, Ayon-Beato:2015qfa, Giardino:2023qlu} cannot occur in 
GR but are typical of scalar--tensor 
gravity.\footnote{Somehow similar solutions, {\em i.e.}, hairy 
Schwarzschild black 
holes in which the scalar field does not gravitate, are known in more 
general Horndeski gravities \cite{Babichev:2012re, Anabalon:2013oea, 
Babichev:2013cya, Charmousis:2014zaa, Kobayashi:2014eva, 
Babichev:2016kdt, BenAchour:2018dap, 
Charmousis:2019vnf, Takahashi:2020hso}.} They have $g_{\mu\nu} = 
\eta_{\mu\nu}$, where $\eta_{\mu\nu}$ 
denotes the Minkowski metric, but $\phi \neq$ constant. In 
other words, 
the gravitational scalar field $\phi$ does not gravitate.

Two types of stealth solutions are most commonly encountered in the 
literature. They have the form 

\begin{enumerate}

\item $g_{\mu\nu} = \eta_{\mu\nu} \quad \text{and} \quad \phi(t) = \phi_0 \, 
\mbox{e}^{\alpha_0 t} $, ~or

\item $g_{\mu\nu} = \eta_{\mu\nu} \quad \text{and} \quad \phi (t)= \phi_0 \,
|t|^{\beta} $,

\end{enumerate}

\noindent where $\phi_0, \alpha_0$, and $\beta$ are constants. Since 
stealth solutions are special cases of FLRW geometries in which the scale 
factor reduces to a constant, they belong to our study. We discuss these 
two cases separately.

\subsection{Stealth solutions $\phi(t) = \phi_0 \, \mbox{e}^{\alpha_0 t}$}

In this case $X = \alpha^2_0 \, 
\phi^2 /2$, yielding
\be
\bar{g}_{\mu\nu} = \eta_{\mu\nu} + F \, \nabla_{\mu} \phi\nabla_{\nu} \phi 
= 
\eta_{\mu\nu} + 
\alpha_0^2 \, F \phi^2 \, \delta_{\mu}^0 \, \delta_{\nu}^0
\ee
and
\be
d\bar{s}^2 = -\left( 1-\alpha_0^2 \, F^2 \phi^2 \right) dt^2 + 
d\vec{x}^{\,2} 
\ee 
where $ F\left( \phi, X \right) = F \left( \phi, \alpha_0^2 \, \phi^2 /2
\right) = F(t)$, which produces
\be
d\bar{s}^2 = -d\tau^2 + d \vec{x}^{\, 2}\,.
\ee
This is another Minkowski line element with time reparametrized. 
We need $ 1 > 
\alpha_0^2 \, \phi^2 \, F(\phi,X) $, or 
\be
F(t) < \frac{1}{\alpha_0^2 \, \phi^2} = 
\frac{ \mbox{e}^{-2\alpha_0t}}{\alpha_0^2 \, \phi_0^2} \,.
\ee

\subsection{The case $\phi(t) = \phi_0 |t|^{\beta}$}

In this second case we have $ \dot{\phi} = \beta \, \phi /t$, 
with 
\be
X =  \frac{\beta^2\phi^2}{2 \,t^2} = \frac{\beta^2 \phi_0^2}{2} \,\,  
t^{2(\beta - 1)} \,, 
\ee
and the disformed line element reads 
\begin{eqnarray}  
 d\bar{s}^2 &=& \left( \eta_{\mu\nu} + F\, \nabla_{\mu} \phi\nabla_{\nu} 
\phi \right)  
dx^{\mu}  dx^{\nu}   
\nonumber\\
&& \nonumber\\
& = & -dt^2 + d\vec{x}^{\, 2} + F\, \dot{\phi}^2 dt^2 = - 
\left( 1-\frac{F\beta^2\phi^2}{t^2}\right)dt^2 + d\vec{x}^{\,2}  
\nonumber\\
&& \nonumber\\
& = & - d\tau^2 + d \vec{x}^{\, 2} \,,
\end{eqnarray}
yielding another Minkowski metric with reparametrized time 
$\tau(t)$.

\section{de Sitter solutions with non--constant scalar field}
\label{sec:7}
\setcounter{equation}{0}

Typical of scalar--tensor gravity is the ``stealth--de Sitter'' 
metric
\be
ds^2 = -dt^2 + a_0^2 \, \mbox{e}^{2H_0 t} d \vec{x}^{\, 2}
\ee
with $ a_0, H_0$ constants, while the scalar field depends on the comoving 
time, 
$\phi = \phi(t)$. 

In GR with a minimally coupled scalar field, de Sitter solutions are 
obtained only for a {\em constant} scalar field $\phi=\phi_0$, and they 
are equilibrium points of the dynamical system formed by the 
Einstein--Friedmann equations describing FLRW cosmology 
\cite{EllisWainwright1997,Coley:2003mj,Faraoni:2012bf,Remmen:2013eja}. In 
scalar--tensor 
gravity, where the scalar $\phi$ has gravitational nature and couples 
explicitly to the Ricci scalar, de Sitter solutions with non--constant 
$\phi(t)$ are possible, but they are not fixed points of the phase space 
of FLRW cosmology.

For de Sitter spacetimes with non--constant  scalar $\phi(t)$ in 
scalar--tensor gravity, the disformed line element is
\begin{eqnarray}    
 d\bar{s}^2 &=& -dt^2 + a_0^2 \, \mbox{e}^{2H_0t} d\vec{x}^{\, 2} + 
F\dot{\phi}^2 dt^2 
\nonumber\\
&& \nonumber\\
& = & - \left( 1-F\dot{\phi}^2 \right)dt^2 + a_0^2 \, \mbox{e}^{2H_0t} 
d\vec{x}^{\, 2} \,,
\end{eqnarray}
or 
\be
d\bar{s}^2 = -d\tau^2 + a_0^2 \, \mbox{e}^{2H_0 t(\tau)} d \vec{x}^{\, 2} 
\ee 
with
\be
\tau(t) = \int dt \sqrt{1-F\dot{\phi}^2} 
\ee
which, in general, constitutes a non--linear relation between $t$ 
and $\tau$. A disformal transformation maps a stealth--de Sitter geometry 
into a less symmetric FLRW geometry with  
\be
d\bar{s}^2 = -d\tau^2 + \bar{a}^2(\tau) \, d \vec{x}^{\, 2} \,, \quad 
\bar{a}(\tau) = a_0 \, \mbox{e}^{H_0 t(\tau)} \,.
\ee
This solution is a de Sitter geometry only if 
$F\dot{\phi}^2$ is constant, {\em 
i.e.}, $F=\mbox{const.}/X$. 

In the literature there are also stealth solutions with a inhomogenous 
scalar field $\phi = \phi \left( t, \vec{x} \right)$ and $g_{\mu\nu} = 
\eta_{\mu\nu}$ ({\em e.g.}, \cite{Ayon-Beato:2005yoq}), leading to
\be
\nabla_{\mu}\phi = \dot{\phi} \, \delta_{\mu}^0 + \phi_i \, 
\delta_{\mu}^i
\ee
and 
\begin{eqnarray}  
\nabla_{\mu}\phi \nabla_{\nu}\phi &=& \left( \dot{\phi} \,  
\delta_{\mu}^0 +  \phi_i \, \delta_{\mu}^i \right) \left( \dot{\phi} \, 
\delta_{\nu}^0 + 
\phi_j 
\, \delta_{\nu}^j \right) \nonumber\\
&& \nonumber\\
& = & \dot{\phi}^2 \, \delta_{\mu}^0 \, \delta_{\nu}^0 + \dot{\phi}  
\left( \phi_j \, \delta_{\mu}^0 \, \delta_{\nu}^j + \phi_i \, 
\delta_{\nu}^0 \,  \delta_{\mu}^i \right)  + \phi_i \, \phi_j \, 
\delta_{\mu}^i \, \delta_{\nu}^j \,, \nonumber\\
&&
\end{eqnarray}
where $\phi_i \equiv \partial\phi /\partial x^i$. 
Additionally, the disformed line element is
\begin{eqnarray}
   d\bar{s}^2 &=& \left( \eta_{\mu\nu} + F\, \nabla_{\mu}\phi 
\nabla_{\nu}\phi \right) dx^{\mu} dx^{\nu} \nonumber\\
&& \nonumber\\
& = & -dt^2 + d\vec{x}^{\, 2} + F \left(\dot{\phi}^2 dt^2 + 2 
\dot{\phi}\phi_i 
dt dx^i + \phi_i \phi_j dx^i dx^j \right) \nonumber\\
&&
\end{eqnarray}
 or 
\begin{eqnarray}
d\bar{s}^2 &=& - \left( 1-F\dot{\phi}^2 \right)dt^2 + 2F 
\, \dot{\phi} \, \phi_i \, dt dx^i \nonumber\\
&&\nonumber\\
&\, & + \left( \delta_{ij} + F\phi_i\phi_j \right) dx^i dx^j \,.
\end{eqnarray}
Now 
\be
X = -\frac{1}{2} \left( -\dot{\phi}^2 + \delta^{ij} \, \phi_i \, \phi_j 
\right) = \frac{\dot{\phi}^2}{2} - \frac{(\vec{\nabla}\phi)^2}{2} 
\ee
depends on the spatial coordinates $x^i$. Since $F\left( 
\phi, X 
\right) \dot{\phi}^2$ now depends on $x^i$ as well, it is not possible to 
redefine the time coordinate $\tau$ as we did before 
because $d\tau = \sqrt{ 
1-F\dot{\phi}^2 } \, dt$ is no longer an exact differential. The disformal 
transformation then sends the Minkowski metric into an inhomogenous 
non--stationary geometry.

\section{More general disformal transformations} 
\label{sec:8}
\setcounter{equation}{0}

Instead of pure disformal transformations~(\ref{puredisformal}), one can 
allow for disformal transformations of the more general form 
\cite{BenAchour:2016cay}
\be
g_{\mu\nu} \to \bar{g}_{\mu\nu} = A(\phi,X)g_{\mu\nu} + 
B(\phi,X)\nabla_{\mu} \phi\nabla_{\nu} \phi \,, \label{impuredisformal}
\ee
which we discuss in this section for FLRW and Bianchi geometries. Several, 
but not all, of the results valid for pure disformal transformations still 
hold. 

The inverse of the disformed metric~(\ref{impuredisformal}) is 
\cite{BenAchour:2016cay,Alinea:2020sei}
\be
\bar{g}^{\mu\nu} = \frac{1}{A}\left(g^{\mu\nu} - 
\frac{B}{A-2BX}\nabla^{\mu} \phi\nabla^{\nu} \phi\right) 
\ee
and the invertibility condition is \cite{BenAchour:2016cay,Alinea:2020sei}
\begin{eqnarray}  
& & A \neq 0\,, \quad A+2XA_x - 4X^2B_x \neq 0 \,, \nonumber\\ 
&& \nonumber\\ 
& & A - 2BX \neq 0 \,,\label{urca}
\end{eqnarray}
where the last equation is needed to guarantee that 
\be
\bar{X} = \frac{X}{A-2BX}
\ee
remains well--defined. For  a line element of the form 
\be
ds^2 =-dt^2 +g_{ij}dx^i dx^j
\ee
in the uniform--$\phi$ gauge, we have $X= \dot{\phi^2}/2$, $A(\phi, 
X)=A(t)$, $B(\phi,X)=B(t)$ and the disformed line element is
\begin{eqnarray}
d\bar{s}^2 &=& \bar{g}_{\mu\nu} dx^{\mu} dx^{\nu} = 
\left( Ag_{\mu\nu}+B \, \nabla_{\mu}\phi \nabla_{\nu}\phi \right)
dx^{\mu} dx^{\nu}  \nonumber\\
&&\nonumber\\
&= & -\left( A-B\dot{\phi}^2 \right) dt^2 +A g_{ij} dx^i dx^j\nonumber\\
&&\nonumber\\
& \equiv & -d\tau^2 +\bar{g}_{ij} dx^i dx^j \,,
\end{eqnarray}
where  $\bar{g}_{ij} = A(t) \, g_{ij}$ and
\be
\tau(t)=\int \sqrt{ A(t)-B(t) \dot{\phi}^2 } \, dt \,. 
\ee
This integral is well defined as 
long as $A-B\dot{\phi}^2>0$ since the integrand 
depends only on time and $d\tau$ is an exact differential, and 
$\bar{g}_{ij} = A(t) \, g_{ij}$. For FLRW metrics with $g_{ij}= 
a^2(t) 
\gamma_{ij}(x^k)$, the Hubble function in the 
disformed world is
\begin{eqnarray}
\bar{H} & \equiv & \frac{1}{\bar{a}} \, \frac{d \bar{a}}{d\tau} = 
\frac{1}{\sqrt{A} \, a}\,
\frac{d\left( \sqrt{A} \, a\right)}{dt} \, \frac{dt}{d\tau} \nonumber\\
&&\nonumber\\
&=& \frac{1}{\sqrt{A-B\dot{\phi}^2}} \left( \frac{\dot{A}}{2A}+H \right) 
\,.
\end{eqnarray}
The fixed points $\left( H_0, \phi_0 \right)$  of the 
phase space are mapped 
into new fixed points of the disformed phase space 
\be
\left( H_0 \,, \phi_0 \right) \to \left( \bar{H}_0= 
\frac{ H_0}{\sqrt{A_0}}, \phi_0 \right)
\ee
since $X_0=0$ and $A(\phi, X) = A(\phi_0, 0) \equiv A_0$. 

Setting 
\be
\gamma^2 \equiv A-B\dot{\phi}^2 >0 \,,
\ee
we have $ \bar{X}= X/ \gamma^2 $, 
\be
\bar{u}^{\mu}= \frac{\bar{g}^{\mu\nu}\nabla_{\nu}\phi }{ \sqrt{2\bar{X}} } 
= 
\frac{u^{\mu}}{\gamma} \,,
\ee
while
\be
\bar{u}_{\mu} =\bar{g}_{\mu\nu} \bar{u}^{\nu} = \gamma \, u_{\mu} \,.
\ee
 
The Riemannian 3--metric $h_{\mu\nu}$ on the 3--spaces orthogonal to 
$u^{\mu}$ changes under disformal mappings of the 
type~(\ref{impuredisformal}) according to 
\begin{eqnarray}
\bar{h}_{\mu\nu} & \equiv & \bar{g}_{\mu\nu} +\bar{u}_{\mu} \bar{u}_{\nu} 
= A g_{\mu\nu} + 2X B u_{\mu} u_{\nu} +\gamma^2 u_{\mu} u_{\nu} 
\nonumber\\
&&\nonumber\\
&=&  -A u_{\mu} u_{\nu} +A h_{\mu\nu} +2XB u_{\mu} u_{\nu}+ 
\left(A-2BX\right) u_{\mu} u_{\nu} \nonumber\\
&&\nonumber\\
&=& A h_{\mu\nu} \,.
\end{eqnarray}

When $\nabla^{\mu}\phi$ is timelike and future--oriented, the 
stress-energy tensor of the effective fluid equivalent of $\phi$ in FLRW 
universes is
\begin{eqnarray}
T_{\mu\nu}^{(\phi)} &=& \rho \, u_{\mu} u_{\nu} +P h_{\mu\nu} \nonumber\\
&&\nonumber\\
&=& \rho \, \frac{ \bar{u}_{\mu}}{\gamma} \, \frac{\bar{u}_{\nu}}{\gamma} 
+ \frac{P}{A} \, \bar{h}_{\mu\nu} \nonumber\\
&&\nonumber\\
&=& \bar{\rho} \, \bar{u}_{\mu}\bar{u}_{\nu} +\bar{P} \bar{h}_{\mu\nu} \,,
\end{eqnarray}
where
\be
\bar{\rho} = \frac{\rho}{\gamma^2} \,, \quad \bar{P}=\frac{P}{A} \,.
\ee
The  equation of state parameter of the effective $\phi$--fluid in the 
disformed world is now 
\be
\bar{w} \equiv \frac{ \bar{P}}{\bar{\rho}} = \frac{\gamma^2}{A} \, w \,.
\ee

On the lines of what was already done in Sec.~\ref{sec:5},  
consider a Bianchi 
universe with line element $
ds^2=-dt^2 + g_{ij} dx^i dx^j $ 
in the uniform--$\phi$ gauge in which $\phi=\phi(t)$. This 
universe is mapped into another 
Bianchi universe with line element
\be
d\bar{s}^2=-d\tau^2 + \bar{g}_{ij} dx^i dx^j  \,,
\ee
where $d\tau= \gamma \, dt$ and $\bar{g}_{ij}=A(t) g_{ij}$. 

Unless $A=$~const., stealth solutions 
\be
\left( g_{\mu\nu}, \phi \right)=\left( 
\eta_{\mu\nu}, \phi(t) \right) 
\ee
are not mapped into stealth solutions because 
now the 
disformal transformation  maps the Minkowski line element 
$ds^2=-dt^2+d\vec{x}^2$ 
into
the spatially flat FLRW geometry
\be
d\bar{s}^2 = -d\tau^2 + \bar{a}^2(\tau) \, d\vec{x}^2 \,.
\ee
with rescaled comoving time defined by $d\tau=\gamma \, dt$ and scale 
factor $ \bar{a}(\tau) = \sqrt{A \left( \phi(t), X(t) \right)} \Big|_{t= 
t(\tau) } $.

\section{Higher--order disformal transformations}
\label{sec:9}
\setcounter{equation}{0}

Yet more general disformal transformations have been introduced recently, 
which contain second order derivatives of the scalar 
field instead of first order one \cite{Takahashi:2021ttd}, and even more 
general ones, containing derivatives 
of higher order than second, are  
contemplated  
\cite{Takahashi:2023vva}.  To extend the scope of the previous discussion, 
and of the future search for exact solutions, let us extend the previous 
considerations to second order disformal 
transformations of the form
\be 
\bar{g}_{\mu\nu}= A g_{\mu\nu} +B \nabla_{\mu}\phi \nabla_{\nu}\phi 
+2C \nabla_{(\mu}\phi \nabla_{\nu )} X +D \nabla_{\mu}X \nabla_{\nu}X  
\,,\label{eq:disformed2order} 
\ee
where $A,B,C$ and $D$ are functions of $\phi, X,Y,Z$ with\footnote{For ease of 
comparison with 
Ref.~\cite{Takahashi:2021ttd}, in this section we define $X$ with a different 
sign and normalization than in the previous sections.}   
\begin{eqnarray}
X &\equiv & \nabla_{\mu}\phi \nabla^{\mu}\phi \,,\\
&&\nonumber\\
Y &\equiv & \nabla_{\mu}\phi \nabla^{\mu}X \,,\\
&&\nonumber\\
Z &\equiv & \nabla_{\mu}X \nabla^{\mu}X \,.\nonumber\\
&&
\end{eqnarray}
It is useful to define the quantity 
\begin{eqnarray}
{\cal F} &\equiv & A\left[ A + XB +2YC +ZD \right] \nonumber\\
&&\nonumber\\
&\, & + 
\left( C^2-BD\right) \left( Y^2 -XZ \right)  \,, 
\end{eqnarray}
which will appear in several formulae. Invertibility of the disformal 
transformation corresponds to 
\cite{Takahashi:2021ttd}
\be
A\neq 0 \,, \quad {\cal F}\neq 0 \,, \quad \bar{X}_X\neq 0 \,, \quad \bar{X}_Y=\bar{X}_Z=0 
\,, 
\ee
\be
\left| \frac{ \partial \left( \bar{Y}, \bar{Z} \right)}{\partial \left( Y,Z \right)} \right| 
\neq 0 \,,
\ee
where \cite{Takahashi:2021ttd}
\begin{eqnarray}
\bar{X} & \equiv & \bar{g}^{\mu\nu} \nabla_{\mu} \phi \nabla_{\nu} \phi
= \frac{XA-D\left( Y^2 -XZ\right)}{ {\cal F}}  \,,\\
&&\nonumber\\
\bar{Y} & \equiv & \bar{g}^{\mu\nu} \nabla_{\mu} \phi \nabla_{\nu} X 
= \bar{X}_X \, \frac{YA+C \left( Y^2 -XZ\right)}{ {\cal F}} 
+ \bar{X}_{\phi}\bar{X} \,,\nonumber\\
&&\\
\bar{Z} & \equiv & \bar{g}^{\mu\nu} \nabla_{\mu} X \nabla_{\nu} X 
= \bar{X}_X^2 \, \frac{ZA -B \left( Y^2 -XZ\right)}{ {\cal F}} 
+ 2\bar{X}_{\phi}\bar{Y} \nonumber\\
&&\nonumber\\
&\, & - \bar{X}_{\phi}^2 \bar{X} \,,
\end{eqnarray}
where 
$\bar{X}_{\phi} \equiv \partial \bar{X}/\partial_{\phi}$, 
$\bar{X}_X \equiv \partial \bar{X}/\partial_X$. The inverse of the 
disformed metric~(\ref{eq:disformed2order}) is
\begin{eqnarray}
\bar{g}^{\mu\nu} &=& 
\frac{1}{A} \left\{ g^{\mu\nu} 
-\frac{ \left[ AB-Z\left( C^2-BD\right)\right]}{ {\cal F}} \, 
\nabla^{\mu}\phi \nabla^{\nu}\phi 
\right.\nonumber\\
&&\nonumber\\
&\, & \left. -2 \, \frac{\left[ AC+Y\left( C^2-BD\right)\right]}{ {\cal 
F}} \,  \nabla^{(\mu}\phi \nabla^{\nu )} X \right.\nonumber\\
&&\nonumber\\
&\, & \left. 
-\frac{\left[ AD-X\left( C^2-BD\right) \right]}{ {\cal F}} \, 
\nabla^{\mu}X \nabla^{\nu}X \right\} \,.
\end{eqnarray}
Let us specialize these formulae to the line element of interest in cosmology
\be
ds^2 =-dt^2 +g_{ij} dx^i dx^j 
\ee
in the uniform $\phi$-gauge in which $\phi=\phi(t)$. It follows 
immediately that
\begin{eqnarray}
X &=& -\dot{\phi}^2 \,,\quad Y=2\dot{\phi}^2\ddot{\phi} \,, \quad
Z = -4\dot{\phi}^2 \ddot{\phi}^2 \,,
\end{eqnarray}
\be
{\cal F} =  A\left( A-B\dot{\phi}^2 +4C\dot{\phi}^2 \ddot{\phi} -4 D
\dot{\phi}^2 \ddot{\phi}^2 \right) \,,
\ee
yielding
\begin{eqnarray}
d\bar{s}^2 &=&  \bar{g}_{\mu\nu}dx^{\mu} dx^{\nu} \nonumber\\
&&\nonumber\\
&=& -\left( A-B\dot{\phi}^2 +4C\dot{\phi}^2 \ddot{\phi} -4 D 
\dot{\phi}^2 
\ddot{\phi}^2 \right)  dt^2 
+Ag_{ij}dx^i dx^j \nonumber\\
&&\nonumber\\
&=& -\frac{ {\cal F} }{A} \, dt^2 +Ag_{ij}dx^i dx^j \,. \label{urcaurca}
\end{eqnarray} 
To preserve the metric signature, it must be $A>0$ and ${\cal F}>0$. Since 
$A, B, C, D$ and $\phi$ and its derivatives depend only of time, $d\tau 
\equiv \sqrt{ \frac{ {\cal F} }{A} } \, dt$ is an exact differential and 
the time coordinate $\tau \equiv \int \sqrt{ \frac{ {\cal F} }{A} } \, dt$ 
is well defined. Using $\tau$, one rewrites the line 
element~(\ref{urcaurca}) as
\be
d\bar{s}^2 = -d\tau^2 +\bar{A}(\tau)g_{ij}dx^i dx^j \,,
\ee
where $\bar{A}(\tau) = A\left( t ( \tau ) \right)$.   For FLRW metrics 
$g_{ij}=a^2 (t) \gamma_{ij}( x^k) $ and
\be
d\bar{s}^2 = -d\tau^2 +\bar{a}^2(\tau) \gamma_{ij}(x^k) dx^i dx^j 
\ee
with $\bar{a}(\tau) =a\left( t(\tau) \right) \sqrt{ A\left( 
t(\tau)\right) } $. The Hubble function in the disformed world is 
\be
\bar{H} \equiv \frac{1}{\bar{a} } \, \frac{ d\bar{a} }{d\tau} = 
\frac{1}{\sqrt{A} \, a } \, \frac{ d\left( \sqrt{A} \, a \right)}{dt} \, 
\frac{dt}{d\tau} 
= \sqrt{ \frac{A}{ {\cal F} } } \left( \frac{\dot{A}}{2A} 
+H \right) \,.
\ee 
Here  $Y^2-XZ=0$ and ${\cal F}$ reduces to
\be
{\cal F}= A \left[ A-B\dot{\phi}^2 + 4\dot{\phi}^2 \ddot{\phi} \left( 
C-D\ddot{\phi} \right) \right] \,,
\ee 
while 
\be
\bar{X} = -\frac{A\dot{\phi}^2}{ {\cal F} } \,.
\ee

Assuming $\nabla^{\mu}\phi$ to be timelike and future--oriented, the 
four--velocities of 
the $\phi$ fluid before and after the disformal transformation are 
\be
u^{\mu} \equiv \frac{ g^{\mu\nu} \nabla_{\nu}\phi}{\sqrt{-X}} = 
- \mbox{sign}\left( 
\dot{\phi} \right) {\delta^{\mu}}_0 
\ee 
and (see Appendix~\ref{appendix:A})
\begin{eqnarray}
\bar{u}^{\mu} & \equiv & \frac{  \bar{g}^{\mu\nu} 
\nabla_{\nu}\phi}{\sqrt{-\bar{X} }} 
\nonumber\\
&&\nonumber\\ 
&=&
\left\{ \sqrt{ \frac{ {\cal F} }{A^3} } +\frac{ \dot{\phi}^2}{ \sqrt{ {\cal F} 
A^3} } \left[ 
A\left( B-2C\ddot{\phi}  \right) \right.\right.\nonumber\\
&&\nonumber\\
&\, & \left. \left. +2 \ddot{\phi} A \left( 2D 
\ddot{\phi} -C \right) 
\right] \right\} u^{\mu} \,,
\end{eqnarray}
that is, $\bar{u}^{\mu}$ is parallel to $u^{\mu}$. Then, by comparing 
Eq.~(\ref{urcaurca}) 
with 
\be
d\bar{s}^2 = \bar{g}_{\mu\nu} dx^{\mu} dx^{\nu} = 
\left( -\bar{u}_{\mu} \bar{u}_{\nu} +\bar{h}_{\mu\nu} \right) dx^{\mu} dx^{\nu} 
\,,
\ee
one obtains immediately
\be
\bar{u}_{\mu} = \sqrt{ \frac{ {\cal F}}{A} } \, u_{\mu} \,,\quad \bar{h}_{\mu\nu}= A 
\, h_{\mu\nu} \,.
\ee 
The fluid sourcing the homogeneous cosmological spacetime has the form
\begin{eqnarray} 
T_{\mu\nu}^{(\phi)} &=& \rho u_{\mu} u_{\nu} +P h_{\mu\nu} = \rho \, \frac{A}{ {\cal F}} 
\, \bar{u}_{\mu} \bar{u}_{\nu} + P \, \frac{ \bar{h}_{\mu\nu}} {A} \nonumber\\
&&\nonumber\\
&=&  \bar{\rho} \, \bar{u}_{\mu} \, \bar{u}_{\nu} + \bar{P} \, \bar{h}_{\mu\nu} \,,
\end{eqnarray}
where 
\be
\bar{\rho} = \frac{\rho A}{ {\cal F}} \,, \quad \bar{P}=\frac{P}{A} \,.
\ee
If $ w\equiv P/\rho$ is the equation of state parameter of the fluid before the 
transformation, its disformed cousin is 
\be
\bar{w} \equiv \frac{ \bar{P} }{\bar{\rho} } = \frac{ {\cal F} }{A^2} \, w \,.
\ee

When it acts on Bianchi universes with line element
\be 
ds^2 = -dt^2 +g_{ij}\left( t, x^i\right) dx^i dx^j
\ee
in the uniform--$\phi$ gauge, the second order disformal 
transformation~(\ref{eq:disformed2order}) produces the new line eleemnt
\be
d\bar{s}^2 = -\frac{ {\cal F}}{A} \, dt^2 + A g_{ij} dx^i dx^j \,.
\ee
Since $ {\cal F}/A =  A-B\dot{\phi}^2 +4C \dot{\phi}^2 \ddot{\phi} 
-4D \dot{\phi}^2 \ddot{\phi}^2 $ depends only on the time $t$, one can 
introduce the new time coordinate defined by $d\tau \equiv \sqrt{ 
\frac{{\cal F}}{A} } \, dt $ (an exact differential) to write
\be
d\bar{s}^2 = - d\tau^2 + \bar{g}_{ij}\left( \tau, x^k \right) dx^i dx^j 
\ee
with $\bar{g}_{ij}\left( \tau, x^k \right) =A( t(\tau)) \, g_{ij}\left( 
t(\tau) , x^k \right)$, which is again a Bianchi geometry with the same 
symmetries of the seed metric $g_{\mu\nu}$, in the uniform--$\phi$ gauge. 

Finally, under the disformal transformation~(\ref{eq:disformed2order}), 
stealth solutions $\Bigg( g_{\mu\nu} \,, \phi(t) \Bigg)= \Bigg( 
\eta_{\mu\nu} \,, \phi (t) \Bigg) $ of the 
field 
equations become again the spatially flat FLRW geometries
\be
d\bar{s}^2 = - d\tau^2 + A\left( t(\tau) \right) d\vec{x}^2 \,,\quad 
\phi=\phi(t)  
\ee
with the uniform--$\phi$ gauge coinciding with the comoving gauge.

\section{Conclusions}
\label{sec:10}
\setcounter{equation}{0}

Disformal transformations have been introduced in gravity long ago 
\cite{Bekenstein:1992pj}, but their use has been greatly revamped only 
recently in the context of Horndeski and DHOST scalar--tensor gravity. 
Pure disformal transformations of FLRW spaces have been studied in 
\cite{Domenech:2015hka}, whose discussion we expand here.  We have 
discussed disformal transformations of cosmological spaces, making 
explicit the $3+1$ splitting, which is essential for the understanding of 
the dissipative 
fluid equivalent of the scalar field $\phi$ of scalar tensor--gravity. The 
latter is well--defined only when the gradient $\nabla^{\mu}\phi$  of the 
scalar field is timelike, which is the case of FLRW and Bianchi 
cosmologies. Bianchi universes and the effective dissipative fluid were 
not considered explicitly in the 
literature on disformal transformations. Novel aspects of scalar field 
cosmology under disformal transformations presented here 
include the transformation properties of de Sitter solutions as fixed 
points of the phase space (which have  constant scalar), stealth 
solutions, and de Sitter solutions with non--constant scalar field typical 
of scalar--tensor gravity. We have answered the question of whether these 
solutions are mapped into solutions of the same kind, first considering 
pure disformal transformations, and then more general transformations.

In the phase space of spatially flat FLRW cosmology, which 
is favoured by observations, the physical variables can be chosen 
to be $\left(H, \phi, \dot{\phi} \right)$. Then, 
necessarily, the fixed points of the Einstein--Friedmann dynamical system 
are de Sitter spaces with constant scalar field.  They are invariant under 
pure disformal transformations. These results extend straightforwardly to 
Bianchi universes.

As we have seen in Sec.~\ref{sec:4}, when its gradient 
$\nabla^{\mu} \phi$ is timelike and future--oriented, the gravitational 
scalar $\phi$ is equivalent to a dissipative effective fluid. This 
effective fluid is the basis for the recent formalism dubbed first--order 
thermodynamics of scalar--tensor gravity in which an effective 
``temperature of gravity'' is introduced to describe the deviations of 
gravity from GR, which is  then regarded as the state 
of zero temperature and thermal equilibrium \cite{Faraoni:2021lfc, 
Faraoni:2021jri}. An equation describing the approach to equilibrium is 
also provided \cite{Faraoni:2021lfc, Faraoni:2021jri}, see 
\cite{Giardino:2023ygc} for a review.  In FLRW spaces, the 
analysis of the $3+1$ splitting of FLRW spacetimes  
given explicitly here provides the 
transformation properties of the effective $\phi$--fluid quantities. 
This derivation provides a parallel to well--known tranformation 
formulae of perfect fluids (including scalar field fluids) under {\em 
conformal} transformations.

Stealth solutions and de Sitter solutions with non--constant scalar fields 
are forbidden in GR and are typical of scalar--tensor gravity. They can be 
regarded as degenerate cases of FLRW universes and can, therefore, be 
analyzed in the same way, as done here in Secs.~\ref{sec:6} 
and~\ref{sec:7}.  

Finally, most of the  results  derived here for pure 
disformal transformations survive under more general ({\em 
i.e.}, not ``pure'') disformal transformations of the 
form~(\ref{impuredisformal}). We have extended the study to 
the disformal transformations containing second order derivatives of 
$\phi$ recently introduced in  
\cite{Takahashi:2021ttd,Takahashi:2023vva}.

This study adds to the current knowledge of 
disformal transformations in scalar--tensor theories of gravity, which has 
seen a very significant increase in the last decade  and are useful 
when searching for examples and counterexamples related to disformal 
transformations in cosmology. Furthermore, we have in mind the  
application of disformal transformations to the search for exact solutions 
of the scalar--tensor  field equations which describe black holes and 
other objects embedded in cosmological spacetimes. Such solutions are 
difficult to find in GR and  ``first--generation'' scalar--tensor gravity 
\cite{Faraoni:2015ula} and will be searched for in more general 
scalar--tensor theories 
using disformal tranformatons of GR ``seeds''. The first step to 
understand their properties will be the knowledge of how the FLRW or 
Bianchi ``backgrounds'', in which they are embedded, behave under 
disformal transformations. This first step has been completed here.

\begin{acknowledgments}

This work is supported, in part, by the Natural Sciences \& Engineering 
Research Council of Canada (grant No.~2023--03234 to V.~F.)  and by a 
scholarship from the Studienstiftung des deutschen Volkes of Germany 
(C.~Z.).

\end{acknowledgments}

\begin{appendices} 
\section{Calculation of the disformed four--velocity 
$ \bar{u}^{\mu} $}
\label{appendix:A}
\renewcommand{\theequation}{A.\arabic{equation}}

\begin{widetext}
\begin{eqnarray}
\bar{u}^{\mu} &=&  \frac{ \bar{g}^{\mu\nu}\nabla_{\nu}\phi }{
\sqrt{-\bar{X} } } \nonumber\\
&&\nonumber\\ 
&=& \frac{1}{A} \Bigg\{
g^{\mu\nu} -  \frac{ \left[AB-Z\left( C^2-BD\right)\right]  }{ {\cal F} }    
\, \nabla^{\mu} \phi\nabla^{\nu} \phi
-2\, \frac{  \left[ AC+YZ\left( C^2-BD\right) \right] }{ {\cal F} }    
\, \nabla^{( \mu} \phi\nabla^{\nu)}  X \nonumber\\
&&\nonumber\\
&\, & -\frac{ \left[ AD-X\left( C^2-BD\right) \right]}{ {\cal F}}  \,  
\nabla^{\mu} 
X\nabla^{\nu} X \Bigg\} \frac{\nabla_{\nu}\phi}{ \sqrt{\dot{\phi}^2 
A/{\cal F} } } \Bigg\} \nonumber\\
&&\nonumber\\
&=& \sqrt{ \frac{ {\cal F}}{A^3} } \,  
\frac{ g^{\mu\nu}\nabla_{\nu}\phi 
}{\sqrt{-X}} -
\frac{1}{ |\dot{\phi}| \sqrt{ {\cal F}A^3} } \Bigg\{ 
\left[ AB- Z(C^2-BD)\right] X\nabla^{\mu}\phi
+2\left[ AC+Y(C^2-BD)\right] \frac{\nabla^{\mu}\phi 
\nabla^{\nu}X\nabla_{\nu}\phi + X\nabla^{\mu}X }{2} \nonumber\\
&&\nonumber\\
&\, & +\left[ AD- X(C^2-BD)\right] \nabla^{\mu}X \nabla^{\nu}X 
\nabla_{\nu}\phi 
\Bigg\} \,.
\end{eqnarray}
Using 
\be
\nabla^{\nu}X \nabla_{\nu}\phi = 2\dot{\phi}^2 \ddot{\phi} \,,
\ee
one obtains
\begin{eqnarray}
\bar{u}^{\mu} &=&  \Bigg\{
\sqrt{ \frac{ {\cal F}}{A^3}}\, \frac{1}{|\dot{\phi}|}  +   
\frac{ |\dot{\phi}|}{ \sqrt{ {\cal F}A^3} } \left\{   
\left[ AB+4 \dot{\phi}^2 \ddot{\phi}^2 (C^2-BD)\right] 
-\left[ AC+ 2\dot{\phi}^2 \ddot{\phi} (C^2-BD)\right] (2\ddot{\phi}) 
+4 \ddot{\phi}^2 \left[ AD+\dot{\phi}^2 (C^2-BD)\right] \right.\nonumber\\
&&\nonumber\\
&\, & \left. -2 \ddot{\phi} \left[ AC+ 2\dot{\phi}^2 \ddot{\phi} 
(C^2-BD)\right]\right\} \Bigg\}|\dot{\phi}| \, u^{\mu} \nonumber\\
&&\nonumber\\
&=& \left\{ \sqrt{ \frac{ {\cal F} }{A^3} } +\frac{ \dot{\phi}^2}{  \sqrt{ 
{\cal F} A^3} } \left[ 
A\left( B-2C\ddot{\phi}  \right)  +2 \ddot{\phi} A \left( 2D 
\ddot{\phi} -C \right) 
\right] \right\} u^{\mu} \,.
\end{eqnarray} 
\end{widetext}

\end{appendices}


\end{document}